\documentclass[a4paper]{article}
\usepackage{epsfig,amsmath,amsfonts,amssymb,setspace,multirow,textcomp}
\usepackage[T1]{fontenc}
\makeatletter
\renewcommand{\@evenfoot}{\hfil \thepage \hfil}
\renewcommand{\@oddfoot}{\hfil \thepage \hfil}
\makeatother

\renewenvironment{thebibliography}[1]{\begin{oldthebibliography}{#1}\setlength{\parskip}{0ex}\setlength{\itemsep}{0ex}}{\end{oldthebibliography}}

\begin{document}
\fontsize{11}{11}\selectfont 
\title{Unidentified aerial phenomena I. Observations of events}
\author{B.E.~Zhilyaev, V.\,N.~Petukhov, V.\,M.~Reshetnyk}
%
\date{\vspace*{-6ex}}
\maketitle
\begin{center} {\small $Main \,Astronomical \, Observatory, NAS \,\, of \, Ukraine, Zabalotnoho \,27, 03680, Kyiv, Ukraine$}\\
{\tt zhilyaev@mao.kiev.ua}
\end{center}

\begin{abstract}
NASA commissioned a research team to study Unidentified Aerial Phenomena (UAP), observations of events that cannot scientifically be identified as known natural phenomena. The Main Astronomical Observatory of NAS of Ukraine conducts an independent study of UAP also.
For UAP observations, we used two meteor stations installed in Kyiv and in the Vinarivka village in the south of the Kyiv region. Observations were performed with colour video cameras in the daytime sky.
We have developed a special observation technique, for detecting and evaluating UAP characteristics.
According to our data, there are two types of UAP, which we conventionally call: (1) Cosmics, and (2) Phantoms. We note that Cosmics are luminous objects, brighter than the background of the sky. We call these ships names of birds (swift, falcon, eagle). Phantoms are dark objects, with contrast from several to about 50 per cent.
We present a broad range of UAPs. We see them everywhere. We observe a significant number of objects whose nature is not clear. Flights of single, group and squadrons of the ships were detected, moving at speeds from 3 to 15 degrees per second.
Some bright objects exhibit regular brightness variability in the range of 10 - 20 Hz.
Two-site observations of UAPs at a base of 120 km with two synchronised cameras allowed the detection of a  variable object, at an altitude of 1170 km. It flashes for one hundredth of a second at an average of 20 Hz.
We use colourimetry methods to determine of distance to objects and evaluate their color characteristics.
Objects RGB colors of the Adobe color system had converted to the Johnson BVR astronomical color system using the color corrections.
Phantom shows the colur characteristics inherent in an object with zero albedos. It is a completely black body that does not emit and absorbs all the radiation falling on it. We see an object because it shields radiation due to Rayleigh scattering. An object contrast makes it possible to estimate the distance using colorimetric methods.
Phantoms are observed in the troposphere at distances up to 10 - 12 km. We estimate their size from 3 to 12 meters and speeds up to 15 km/s.


\end{abstract}

\section*{\sc introduction}
\indent \indent The Pentagon is interested in UFOs and created the All-domain Anomaly Resolution Office (AARO). The AARO's mission will be to synchronise the efforts of the Department of Defense and other U.S. federal departments and agencies to detect, identify, and attribute objects in the airspace of military interest associated with threats to air safety and national security. This includes unidentified anomalous, air, space, underwater and trans-medium objects.

NASA will conduct an independent study of unidentified phenomena in the atmosphere. NASA commissions a research team to study Unidentified Aerial Phenomena (UAP) - that is, observations of events that cannot scientifically be identified as known natural phenomena. The agency's independent research group will be led by astrophysicist David Spergel, formerly chairman of the Department of Astrophysics at Princeton University. Daniel Evans, Research Officer at NASA's Science Mission Directorate, will be the NASA official responsible for organizing the study. 

The Main Astronomical Observatory of NAS of Ukraine conducts an independent study of unidentified phenomena in the atmosphere. Our astronomical work is daytime observations of meteors and space invasions. Unidentified anomalous, air, and space objects are deeply concealed phenomena. The main feature of the UAP is its extremely high speed. 

Helmholtz established that the eye does not fix phenomena lasting less than one-tenth of a second. It takes four-tenths of a second to recognize an event. Ordinary photo and video recordings will also not capture the UAP. To detect UAP, you need to fine-tune (tuning) the equipment: shutter speed, frame rate, and dynamic range (14 - 16 stops).

According to our data, there are two types of UAP, which we conventionally call: (1) Cosmics (COS), and (2) Phantoms (PHA). We note that Cosmics are luminous objects, brighter than the background of the sky. We call them names of birds (swift, falcon, eagle). Phantoms are dark objects, with a contrast, according to our data, from 50\% to several per cent. Both types of UAPs exhibit extremely high movement speeds. Their detection is a difficult experimental problem. They are a by-product of our main astronomical work, daytime observations of meteors and space intrusions.

\section*{\sc OBSERVATIONS AND DATA PROCESSING}

For UAP observations, we used two meteor stations installed in Kyiv and in the Vinarivka village in the south of the Kyiv region. The distance between stations is 120 km. The stations are equipped with ASI 178 MC and ASI 294 Pro CCD cameras, and Computar lenses with a focal length of 6 mm. The SharpCap 4.0 program was used for data recording.
Observations of objects were carried out in the daytime sky. The brightness of the sky, depending on the state of the atmosphere and the distance from the Sun, ranges from minus 3 to minus 5 stellar magnitudes per square arc minute.
We have developed a special observation technique, taking into account the high speeds of the observed objects. The exposure time was chosen so that the image of the object did not shift significantly during exposure. The frame rate was chosen to take into account the speed of the object and the field of view of the camera. In practice, the exposure time was less than 1 ms, and the frame rate was no less than 50 Hz.
Frames were recorded in the .ser format with 14 and 16 bits. Violation of these conditions leads to the fact that objects will not be registered during observations.
To determine the coordinates of objects, the cameras were installed in the direction of the zenith or the Moon.

\subsection*{\sc Results}

Fig.1 shows the shoot of ordinary swift objects at a rate of at least 50 frames per second. Two consecutive shots.
The bright objects in Fig. 1 show a constant brightness. Fig. 5 shows an image of an object about 10 pixels in size (about 3 arc minutes), which indicates the final dimensions of the object and a contrast of about 20\%. Fig. 6 shows the color diagram of the object in the RGB filters of the Adobe color system. Object colors can be converted to the Johnson BVR astronomical color system using the color corrections published in \cite{Zhilyaev}.

\begin{figure}[!h]
\centering
\begin{minipage}[t]{.45\linewidth}
\centering
\epsfig{file = 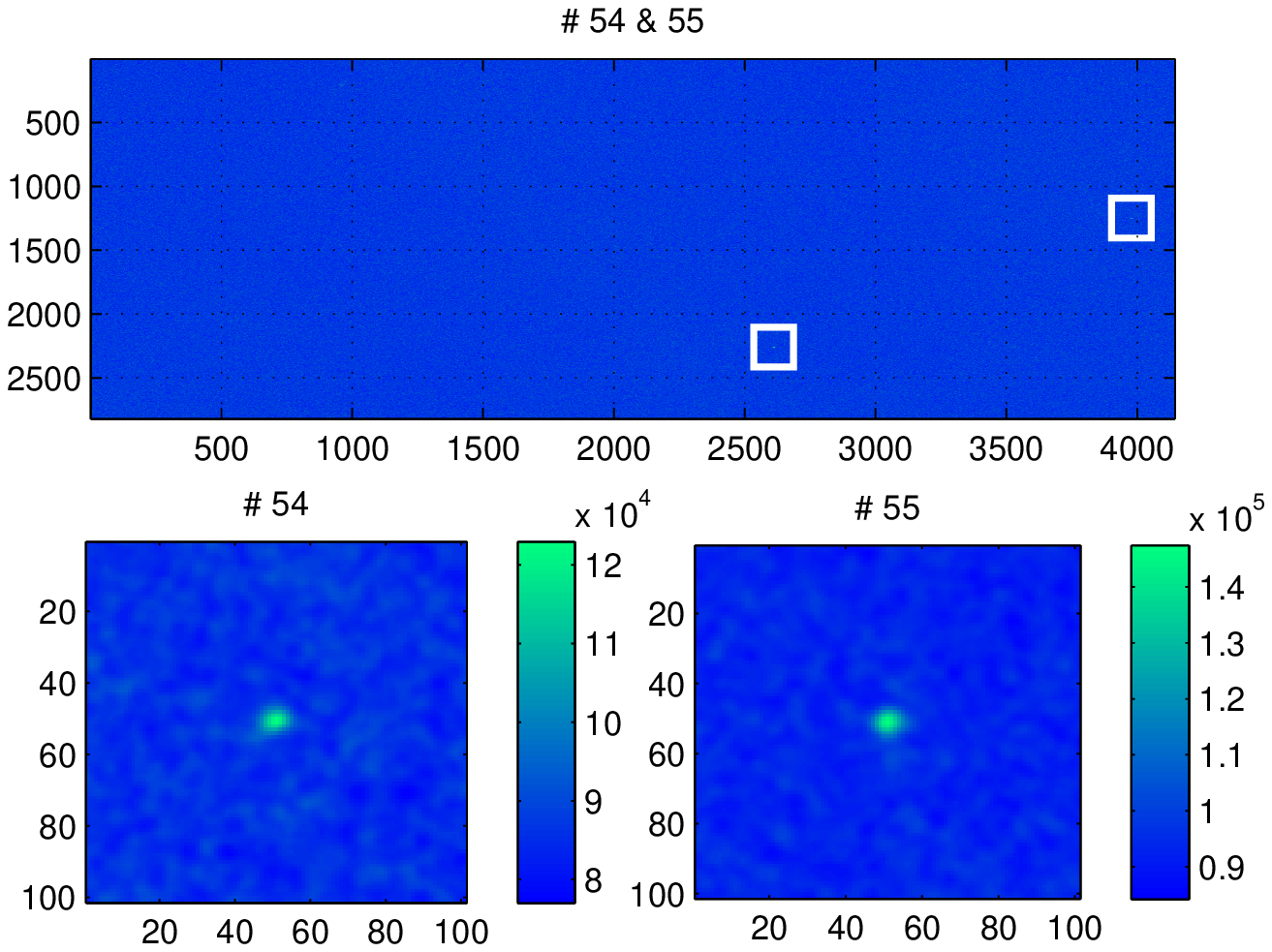,width = 0.9\linewidth} \caption{Two consecutive shots of ordinary swift objects at a rate of at least 50 frames per second.}\label{fig1}
\end{minipage}
\hfill
\begin{minipage}[t]{.45\linewidth} 
\centering
\epsfig{file = 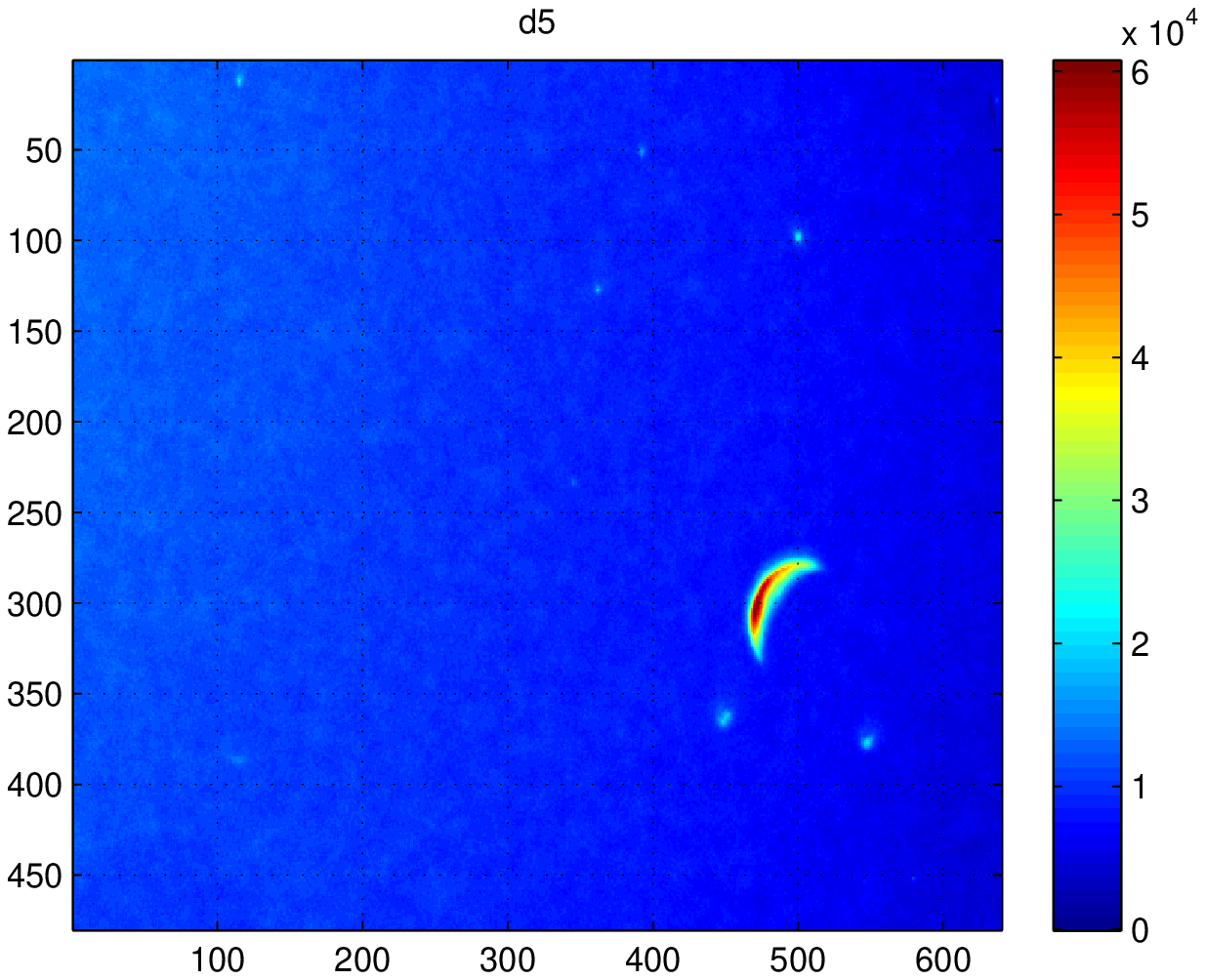,width = 0.9\linewidth} \caption{A group of luminous objects of different brightness against the background of the Moon.}\label{fig2}
\end{minipage}
\end{figure}

\begin{figure}[!h]
\centering
\begin{minipage}[t]{.45\linewidth}
\centering
\epsfig{file = 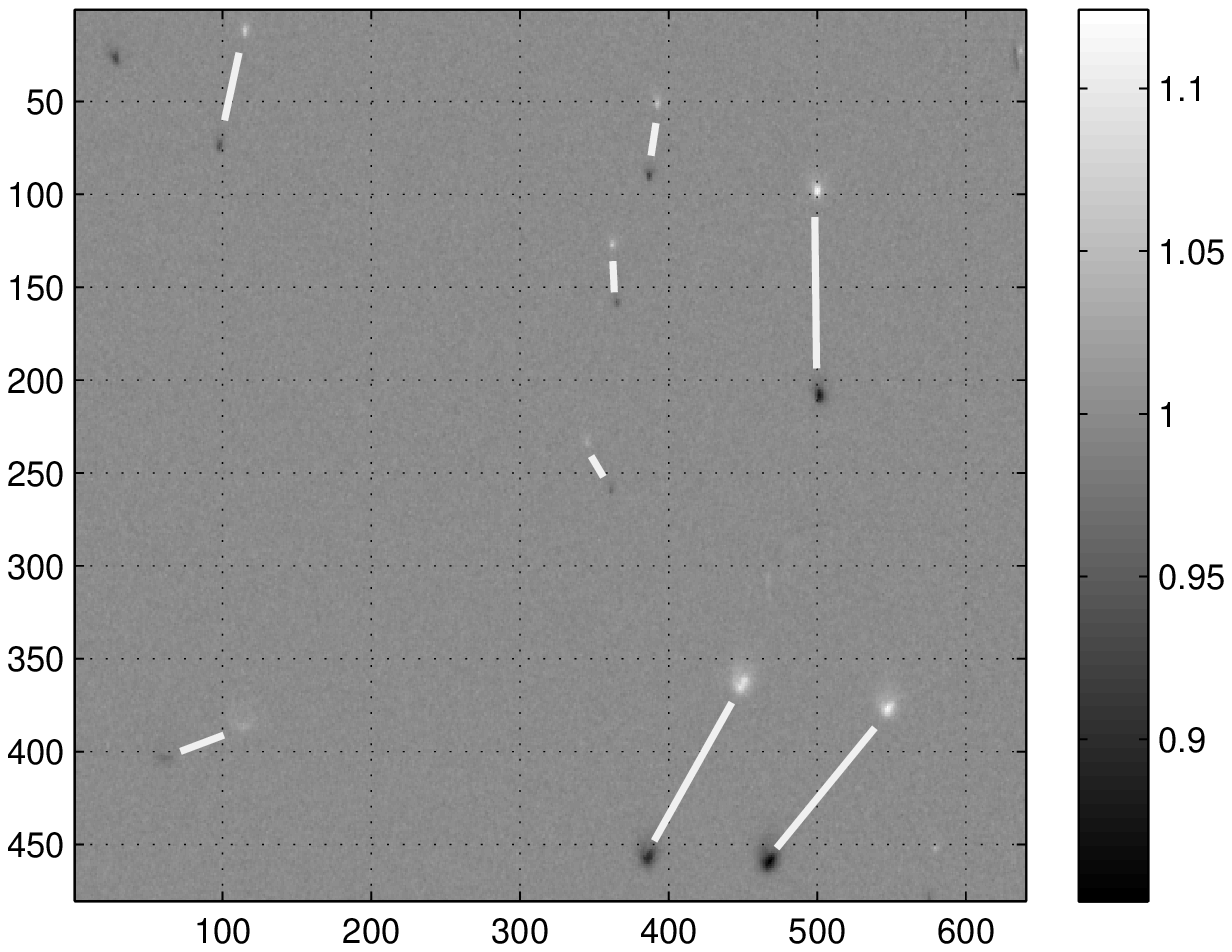,width = 0.9\linewidth} \caption{A composite image with the bright swifts. Segments of straight lines are proportional to transverse speed.}\label{fig1}
\end{minipage}
\hfill
\begin{minipage}[t]{.45\linewidth} 
\centering
\epsfig{file = 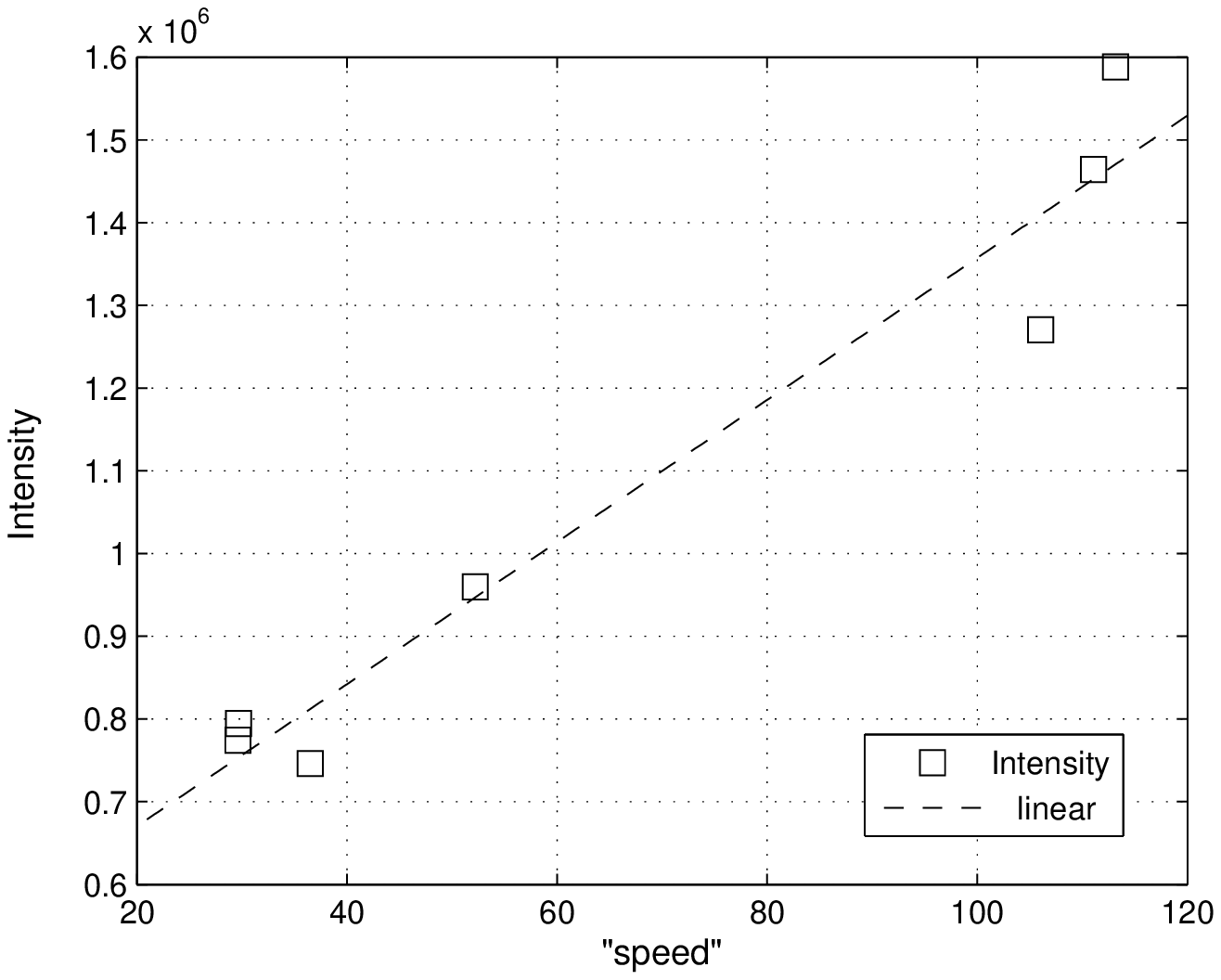,width = 0.9\linewidth} \caption{The intensity of objects versus transverse speed.}\label{fig2}
\end{minipage}
\end{figure}

\begin{equation}\label{}
(B - V)_{J} = (B - G)_{Ad} +0.60 ;\,\, (V - R)_{J} = (G - R)_{Ad} + 0.40
\end{equation}

This makes it possible to compare the colors of the object with the color of the reflected solar radiation. Sun radiation colors (B - V)$_{J}$ =+0.65, (V - R)$_{J}$ =+0.52. The colors of the radiation of the object (B - V)$_{J}$ = +2.86, (V - R)$_{J}$ = +2.88 significantly exceed the colors of the radiation of the Sun.

Fig. 2 shows a group of luminous objects  (flotilla) of class "swift" of different brightness. Objects move at different speeds in different directions. Fig. 3 shows the transversal velocities of the objects. The velocities are represented by segments of straight lines. They obtained from the positions of the objects on two consecutive images. Fig. 4 shows that the "speeds"  correlate with the brightness, namely, the greater the brightness, the greater the speed.

\begin{figure}[!h]
\centering
\begin{minipage}[t]{.45\linewidth}
\centering
\epsfig{file = 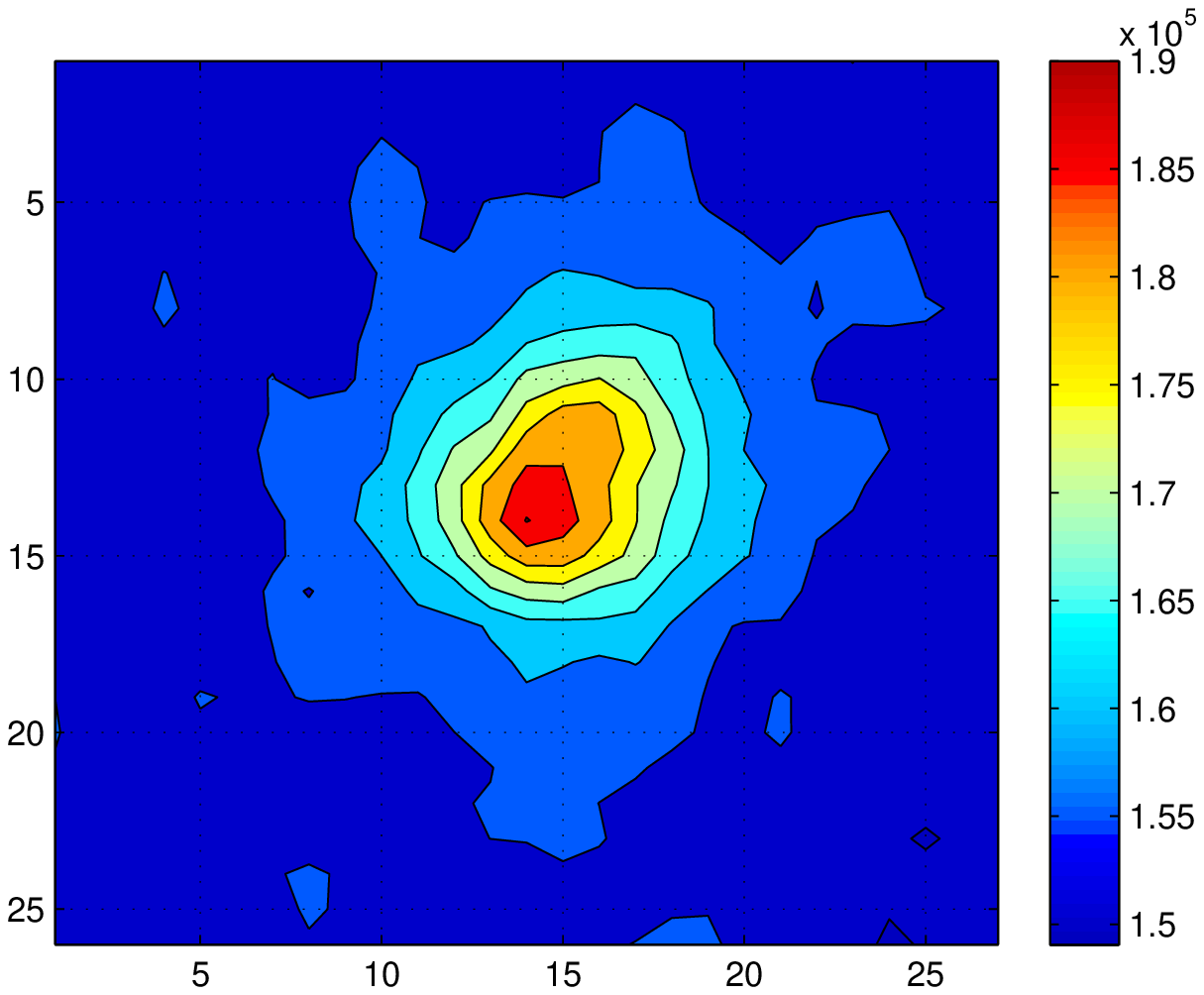,width = 0.9\linewidth} \caption{The swift is the object of end sizes.}\label{fig5}
\end{minipage}
\hfill
\begin{minipage}[t]{.45\linewidth} 
\centering
\epsfig{file = 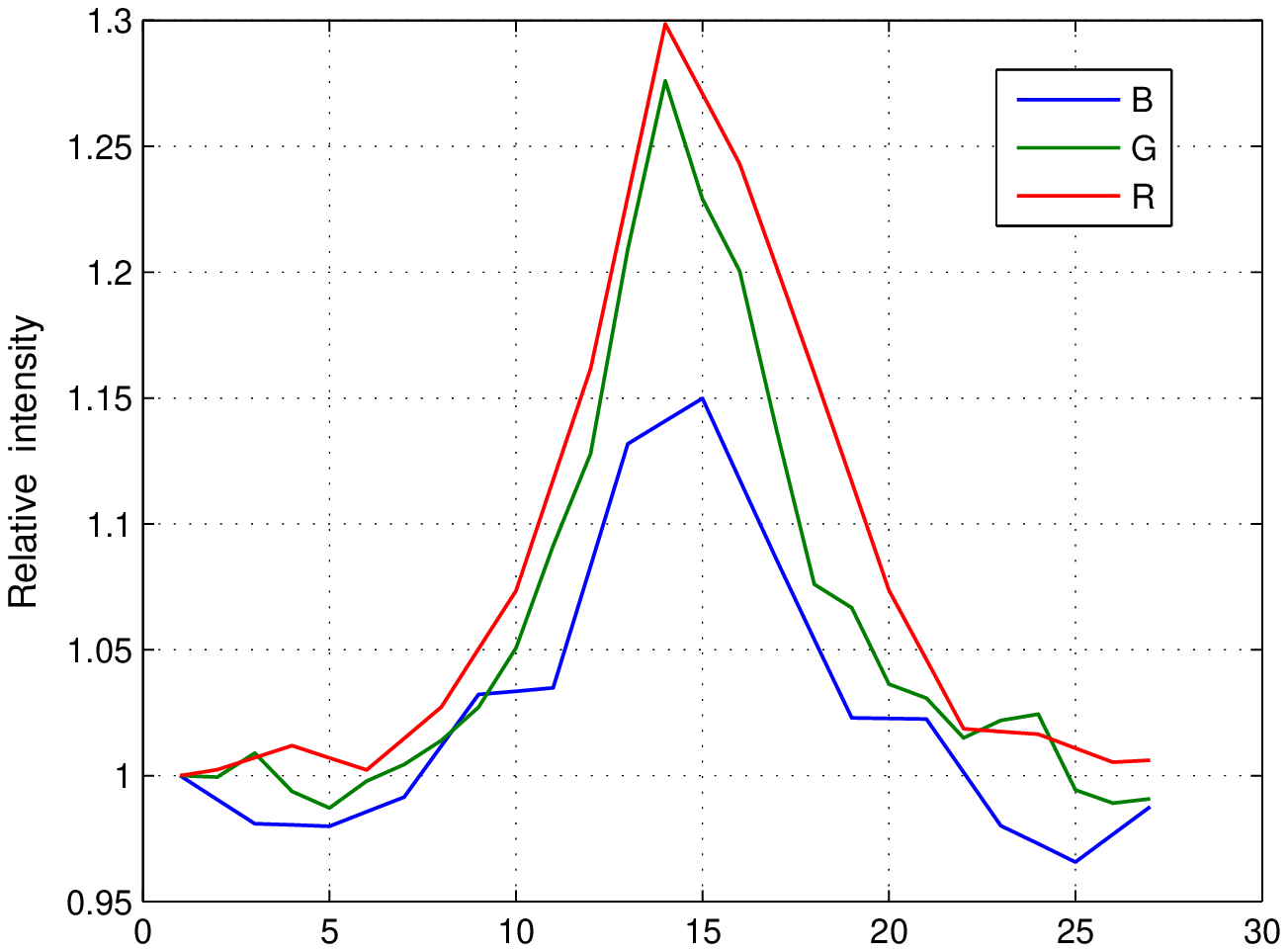,width = 0.9\linewidth} \caption{The RGB emission spectrum of the ordinary swift object.}\label{fig6}
\end{minipage}
\end{figure}

\subsection*{\sc Determination of distance to an object by colorimetry methods}

The colors of the object and the background of the sky make it possible to determine the distance using colorimetric methods. The necessary conditions are (1) Rayleigh scattering as the main source of atmospheric radiation; (2) and the estimated value of the object's albedo. The object partially shields the diffuse sky background and thus becomes visible.
The scattered radiation intensity observed at sea level has the form:

\begin{equation}\label{}
I=I_{0}e^{-\sigma s}
\end{equation}

Here $s$ is the distance to the object, $\sigma $ is the Rayleigh scattering coefficient, and $I_{0}$ is the value of the intensity observed at sea level. The linear Rayleigh scattering coefficient $\sigma $ has the form \cite{Allen}:

\begin{equation}\label{}
\sigma = 3\cdot 10^{18}\cdot \delta \cdot (n-1)^{2}/ \lambda ^{4}/ N
\end{equation}

Here $n$ is the refractive index of air, $ \lambda$ is the wavelength of light in microns, $ \delta$ is the depolarization coefficient equal to 0.97 for the Earth's atmosphere, and $N$ is the number of molecules in 1 cm (Loshmidt number).
Expression (2) can be represented in stellar magnitudes as:

\begin{equation}\label{}
\Delta m=1.086\cdot \sigma\cdot s
\end{equation}

Formally, the magnitude difference $\Delta $m can be considered as a decrease in intensity due to Rayleigh scattering screened by the object against the sky. The value of $\Delta $m per air mass for a clean atmosphere in the visual region (V) is $\Delta$mV $\approx $ 0.20 magnitudes and in the blue region (B) $\Delta$mB $\approx $ 0.34 magnitudes \cite{Allen}.
Thus, by measuring the difference between the stellar magnitudes of an object and the sky background, one can find the magnitude of the air mass before the object. 
We use the approximation of a homogeneous atmosphere for calculations. The homogeneous atmosphere approximation assumes that the entire atmosphere is concentrated in the troposphere (8 - 10 km) and has a constant density. In the approximation of a homogeneous atmosphere by simple algebra, without integration, we get the path length $s$, i.e., distance to the object.
In a real atmosphere, the number of scattering centers (Loshmidt number) at a height of 10 km decreases by a factor of 2.5. When calculating the Rayleigh scattering coefficient in the homogeneous atmosphere approximation in the visual region (V), this introduces an error of about 6\% ($ \sigma$ = 0.251 instead of 0.223 \cite{Allen}.

Figures 7 and 8 show the image and color charts of the phantom object. The object is present in only one frame, which allows us to determine its speed of at least 52 degrees per second, taking into account the angular dimensions of the frame.

Fig. 8 shows the color characteristics inherent in an object with zero albedo. This means that the object is a completely black body that does not emit and absorbs all the radiation falling on it. We see an object only because it shields radiation in the atmosphere due to Rayleigh scattering. An object contrast of about 0.4 makes it possible to estimate the distance to the object as about 5 km. The estimate of the angular velocity given above makes it possible to estimate the linear velocity not less than 7.2 km/s.


\begin{figure}[!h] 
\centering
\begin{minipage}[t]{.45\linewidth}
\centering
\epsfig{file = 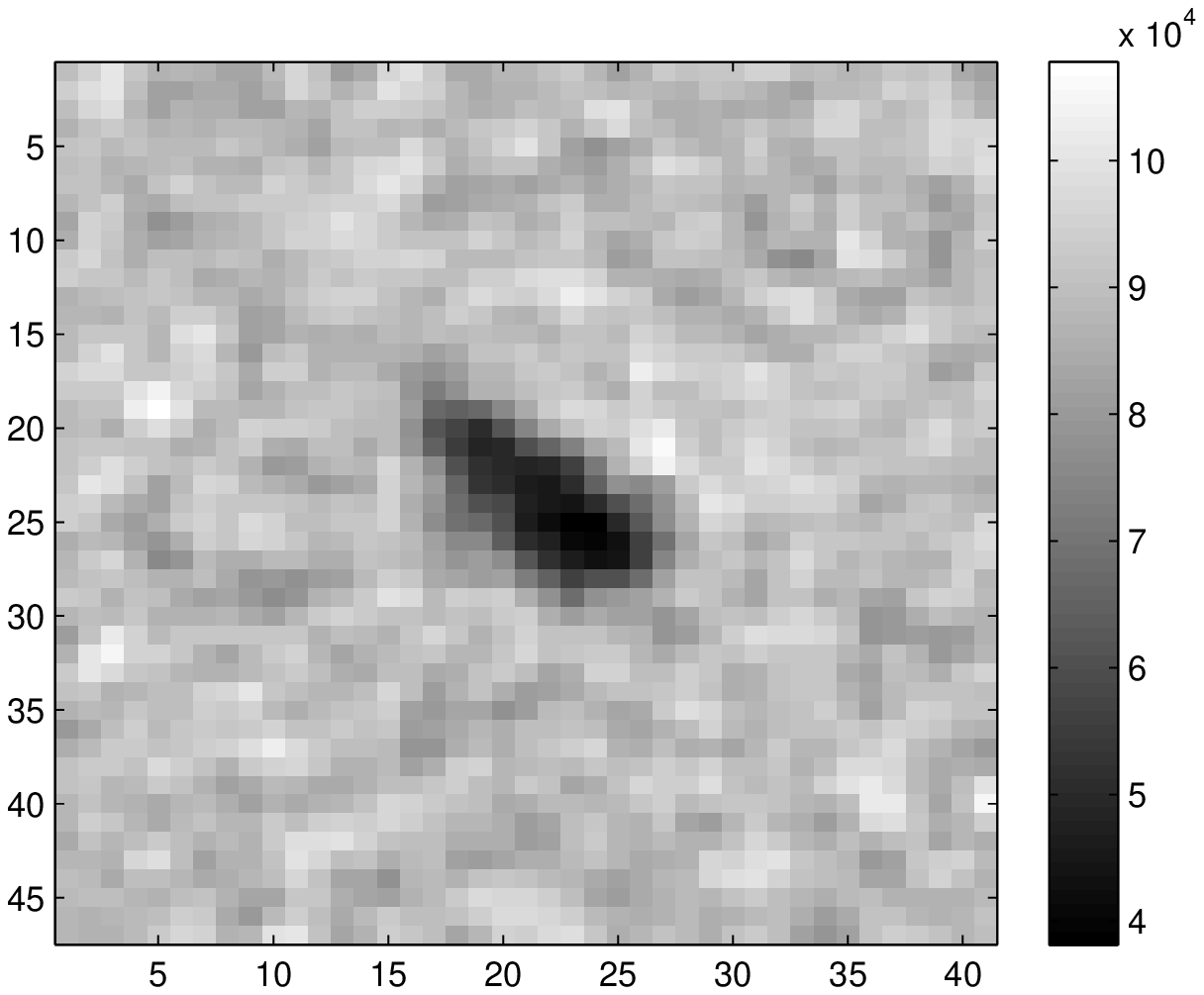,width = 0.9\linewidth} \caption{The image of the phantom object.}\label{fig5}
\end{minipage}
\hfill
\begin{minipage}[t]{.45\linewidth} 
\centering
\epsfig{file = 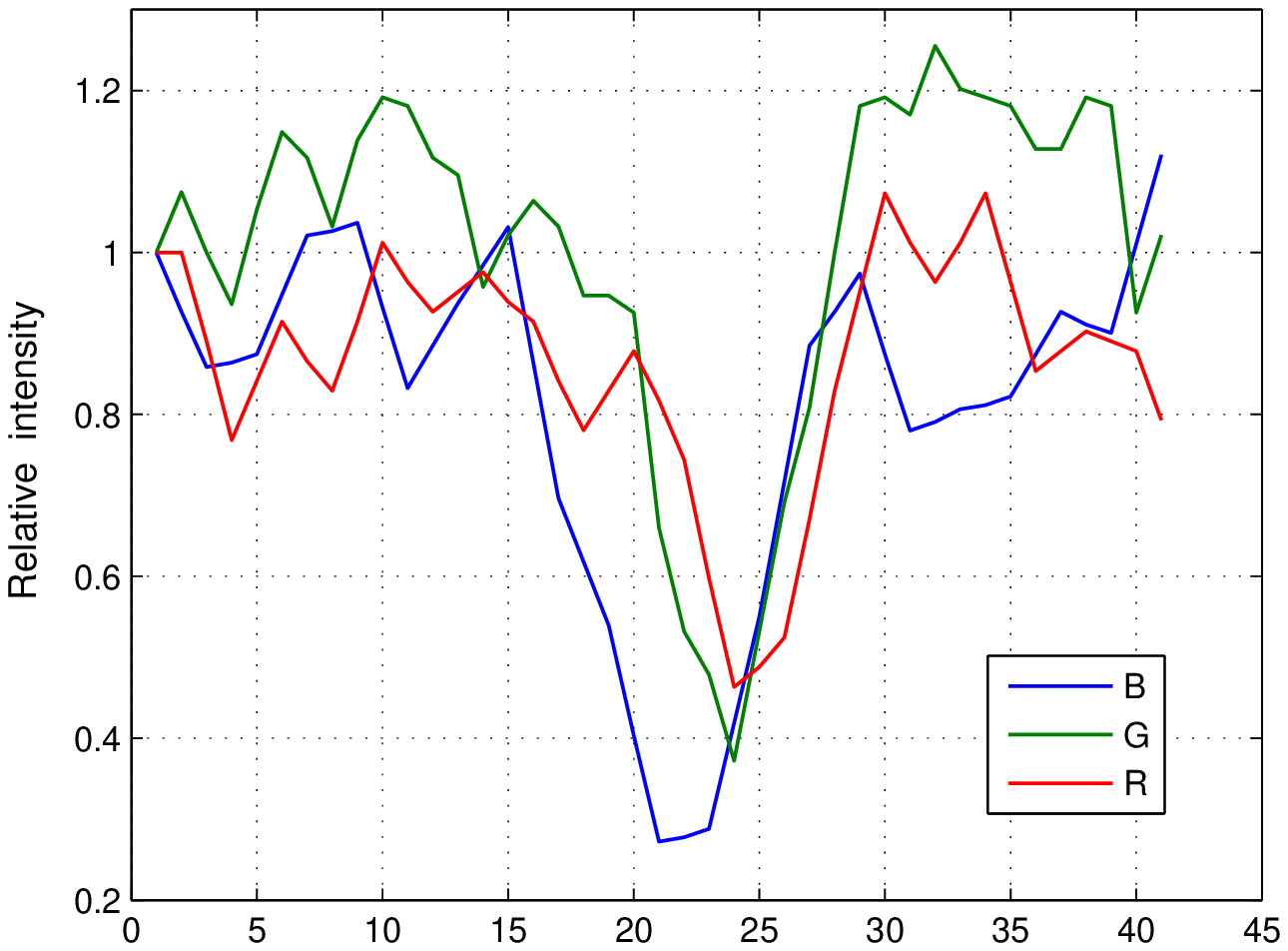,width = 0.9\linewidth} \caption{The color charts of the phantom object.}\label{fig6}
\end{minipage}
\end{figure}

Fig. 9 shows the shoot of another phantom object against the background of the Moon at a rate of at least 50 frames per second. Fig. 10 shows the color diagram of the object and the Moon in the RGB filters of the Adobe color system. Fig. 11 shows an object contrast of about 0.3. It makes it possible to estimate the distance to the object as about 3.5 km. Knowing the distance, we determine the size and speed. Track width is 175 arc seconds, size is 3.0 meters, the track length is 14 meters, exposure time is 1 ms, and speed is 14 km/s.

The color chart in Fig. 10 allows us to evaluate the color characteristics of the Moon and check the calibration of our cameras. The Moon has a color relative to the sky background: B - G = -2.5 log (1.7 / 2.7) = 0.5. We take into account the color correction in the Jhonson B - V system according to [x] due to Rayleigh scattering equal to 0.14 magnitude. Let's get the estimate B - V  of the Moon: B - V = 0.50 + 0.60 - 0.14 = 0.96. The actual color of the Moon is B - V = 0.91 according to \cite{Allen} and differs from our estimate by 0.05 magnitudes within the photometric error.

In Figure 9 we can see a local feature (water tower). The color diagram of the tower in Fig. 12 gives a distance estimate of 0 $\pm $ 1 km. The actual distance is about 300 meters. Thus, colorimetric measurements confirm our estimates.


\begin{figure}[!h] 
\centering
\begin{minipage}[t]{.45\linewidth}
\centering
\epsfig{file = 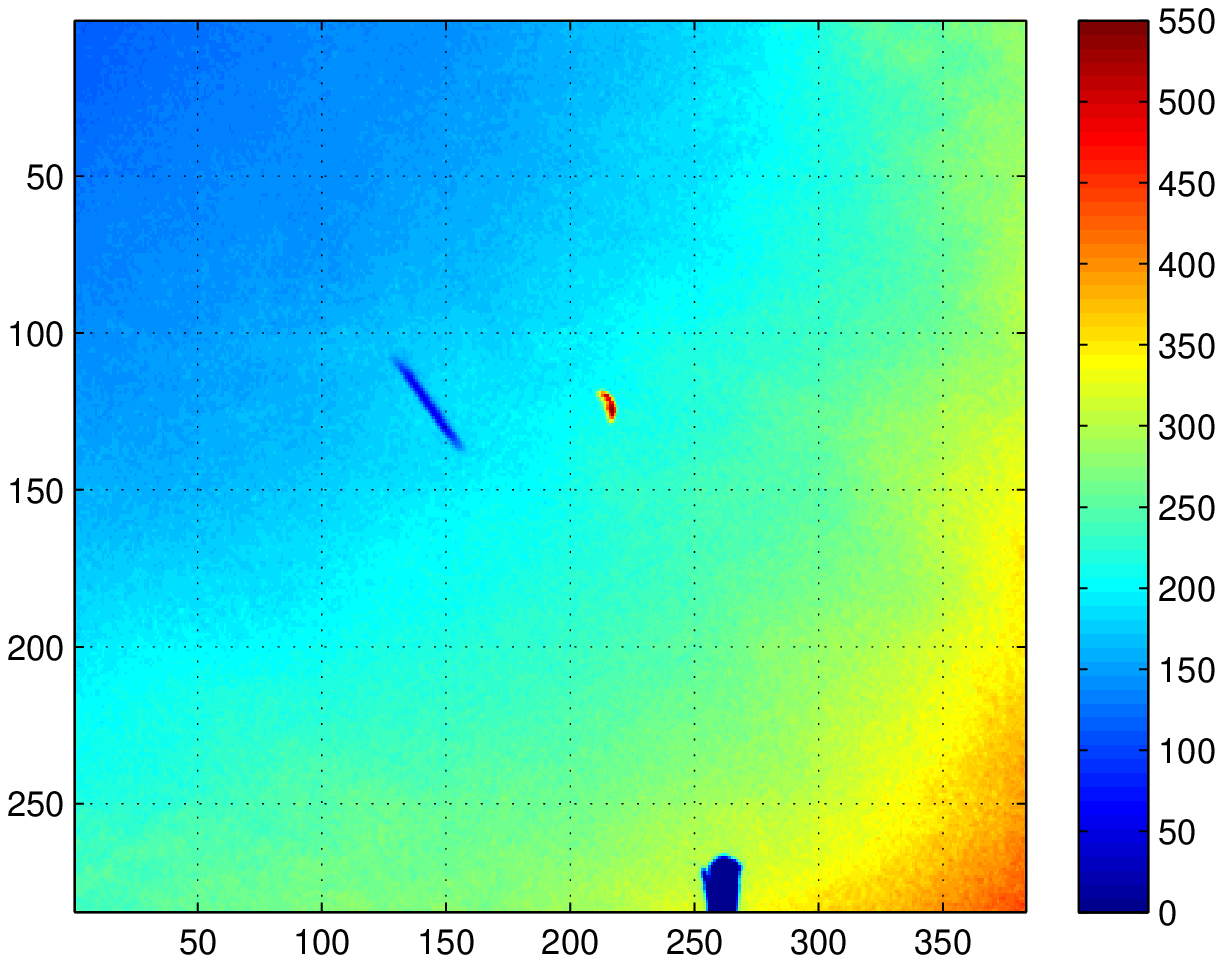,width = 0.9\linewidth} \caption{A phantom object against the background of the Moon.}\label{fig9}
\end{minipage}
\hfill
\begin{minipage}[t]{.45\linewidth}  
\centering
\epsfig{file = 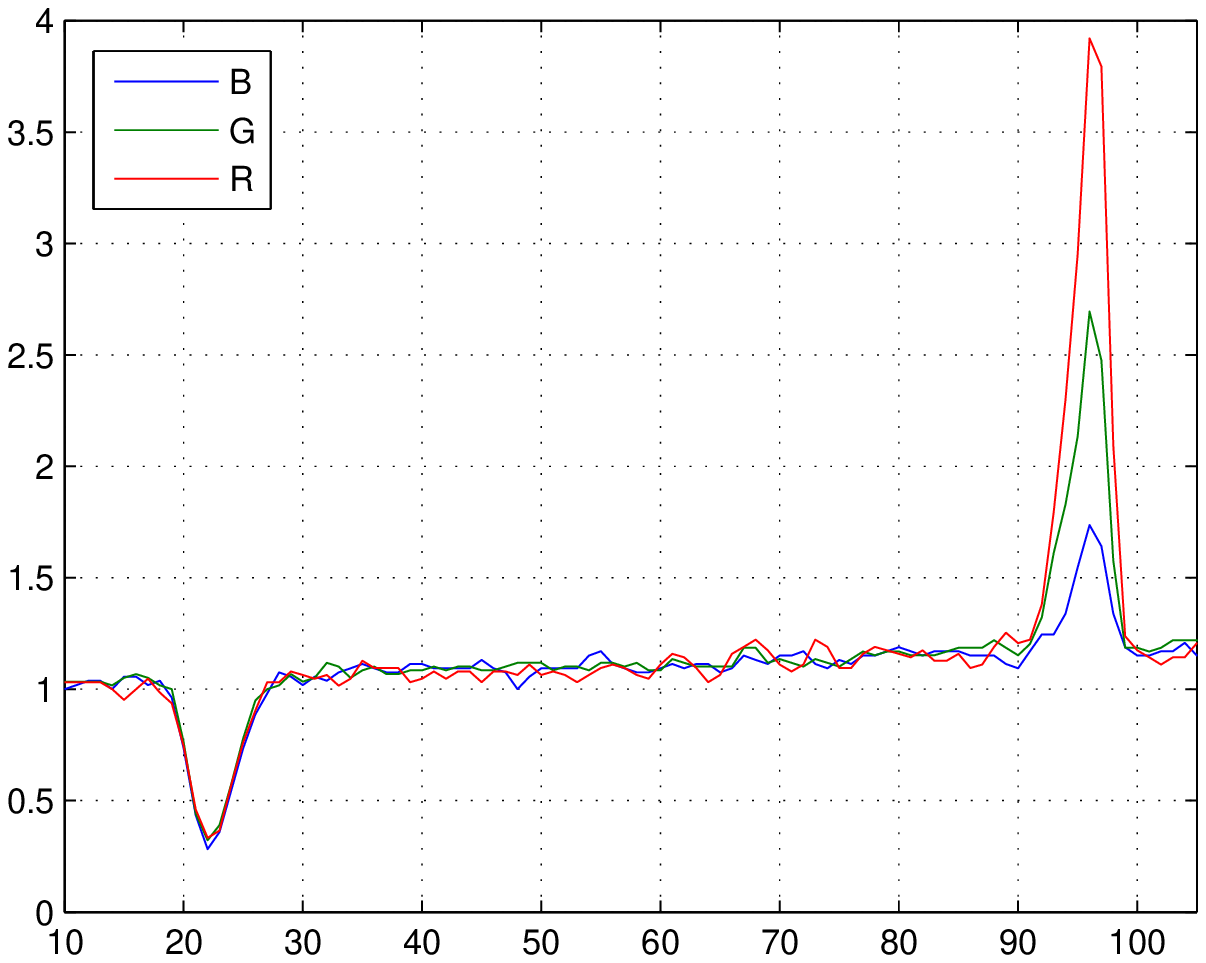,width = 0.9\linewidth} \caption{The color diagram of the objects in the
RGB filters of the Adobe color system.}\label{fig6}
\end{minipage}
\end{figure}


\begin{figure}[!h]  
\centering
\begin{minipage}[t]{.45\linewidth}
\centering
\epsfig{file = 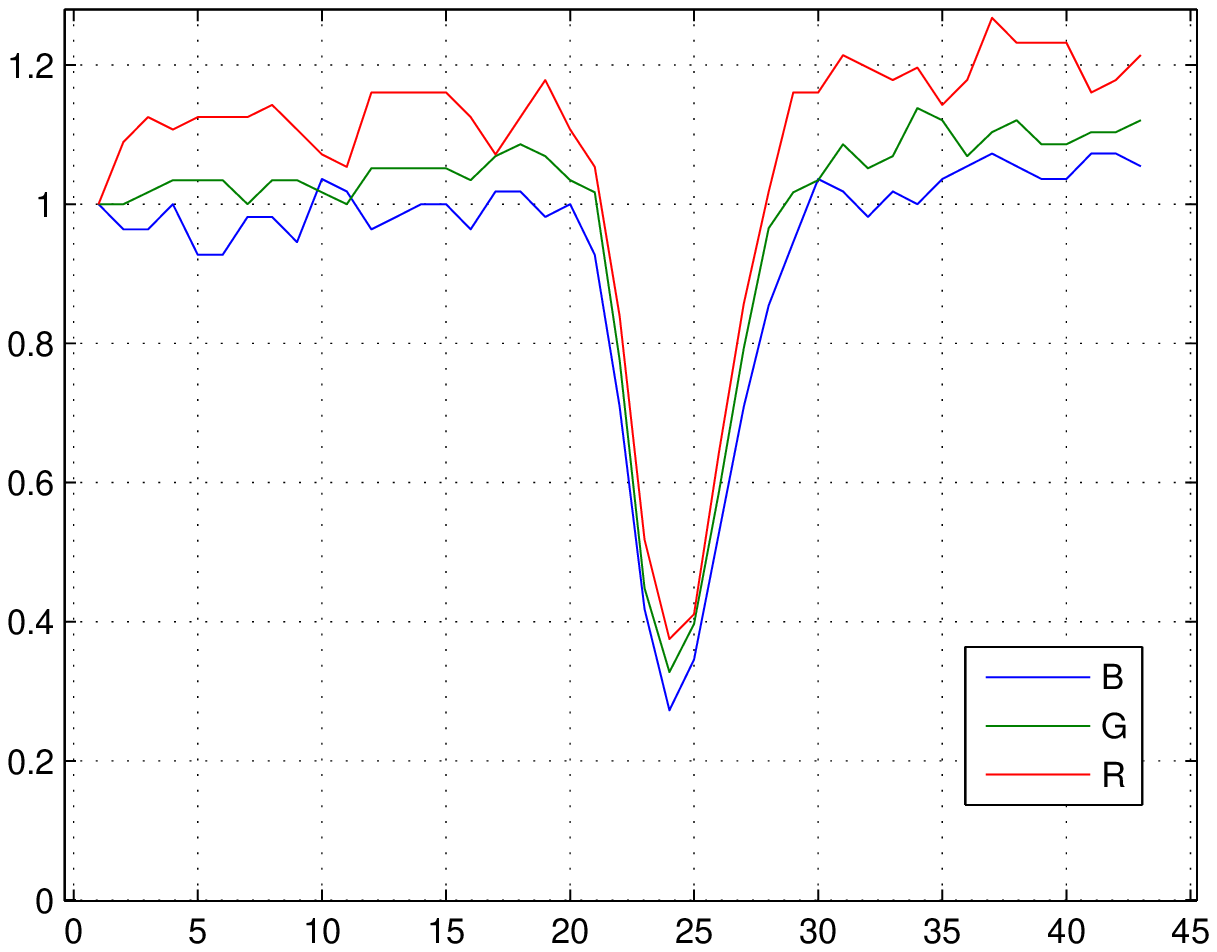,width = 0.9\linewidth} \caption{The color diagram of the phantom.
}\label{fig11}
\end{minipage}
\hfill
\begin{minipage}[t]{.45\linewidth} 
\centering
\epsfig{file = 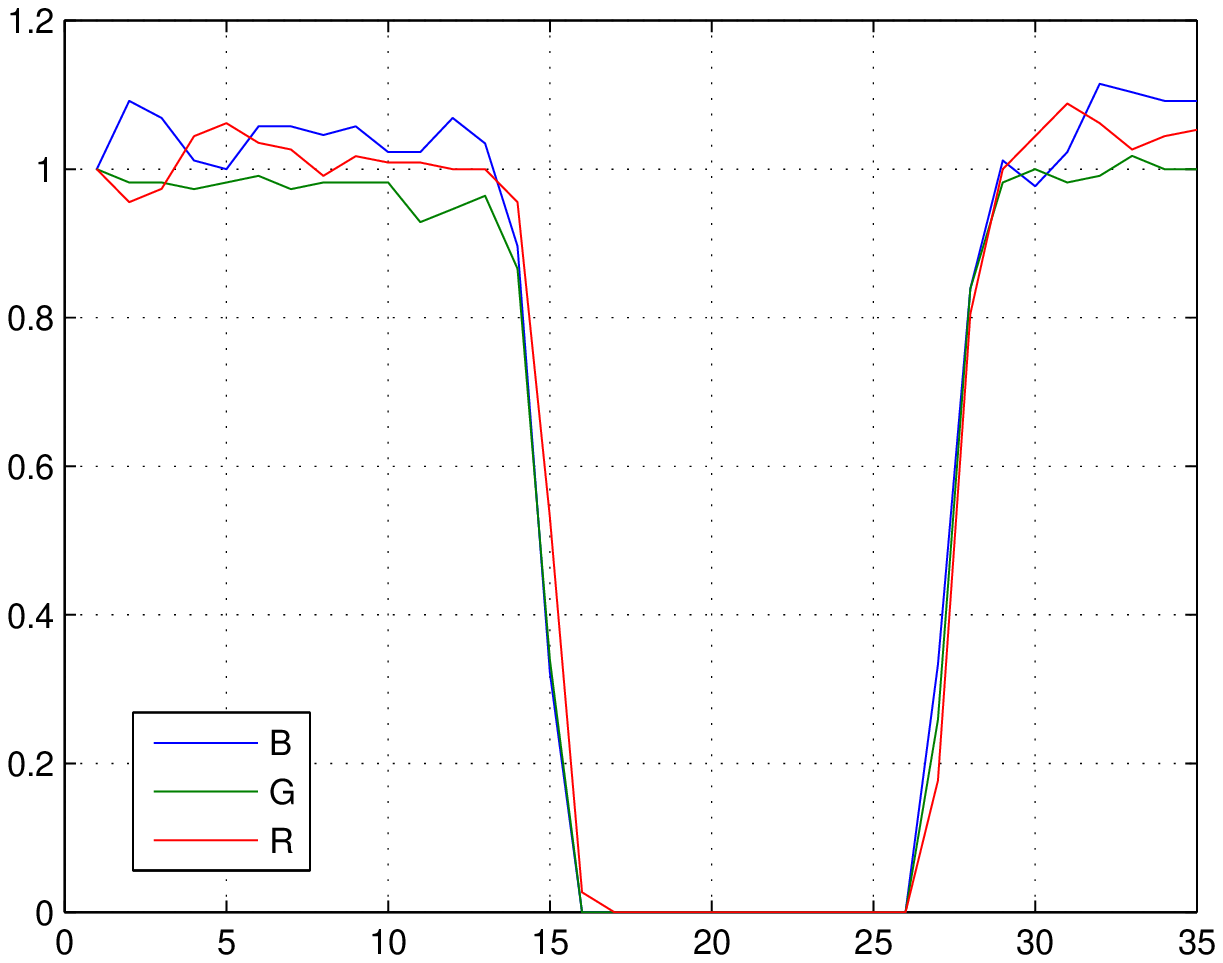,width = 0.9\linewidth} \caption{The color diagram of a local facility.}\label{fig12}
\end{minipage}
\end{figure}

Fig. 13 shows a composite image with the phantom object and bright swifts. Objects move in the same direction, at roughly the same speed. 
Fig. 14 shows an object contrast of about 0.55. This makes it possible to estimate the distance to the object at about 6.0 km. Knowing the distance, we determine the size and speed. The width of the object is 400 arc seconds, the size is about 12.0 meters.
The object crosses the field of view of 3 degrees in 0.18 seconds with a linear speed of about 15 km/s. 

The colors of the swifts radiation  in Fig. 14 significantly differ from the color presented in Fig. 6. 

\begin{figure}[!h] 
\centering
\begin{minipage}[t]{.45\linewidth}
\centering
\epsfig{file = 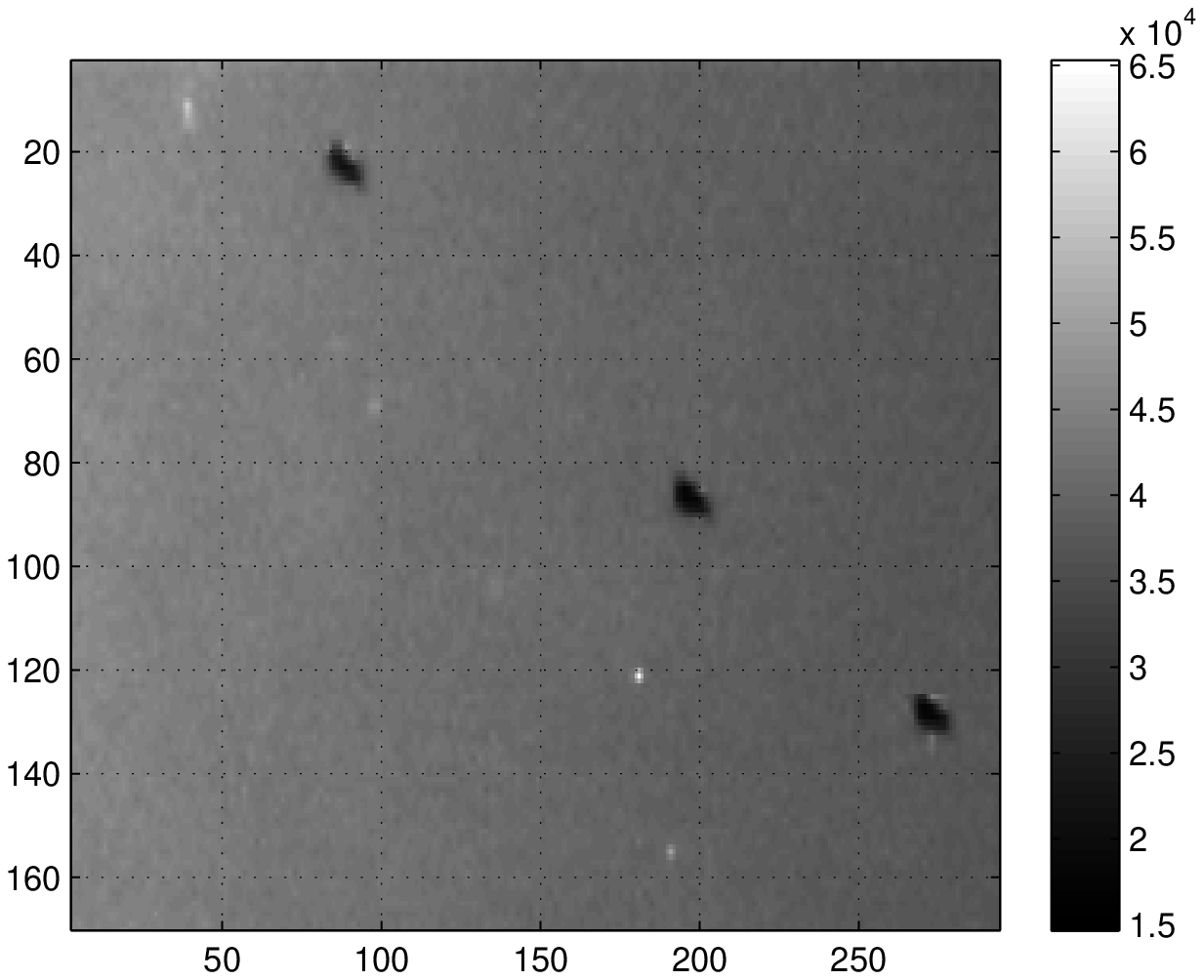,width = 0.9\linewidth} \caption{A composite image with the phantom object and bright swifts.}\label{fig5}
\end{minipage}
\hfill
\begin{minipage}[t]{.45\linewidth} 
\centering
\epsfig{file = 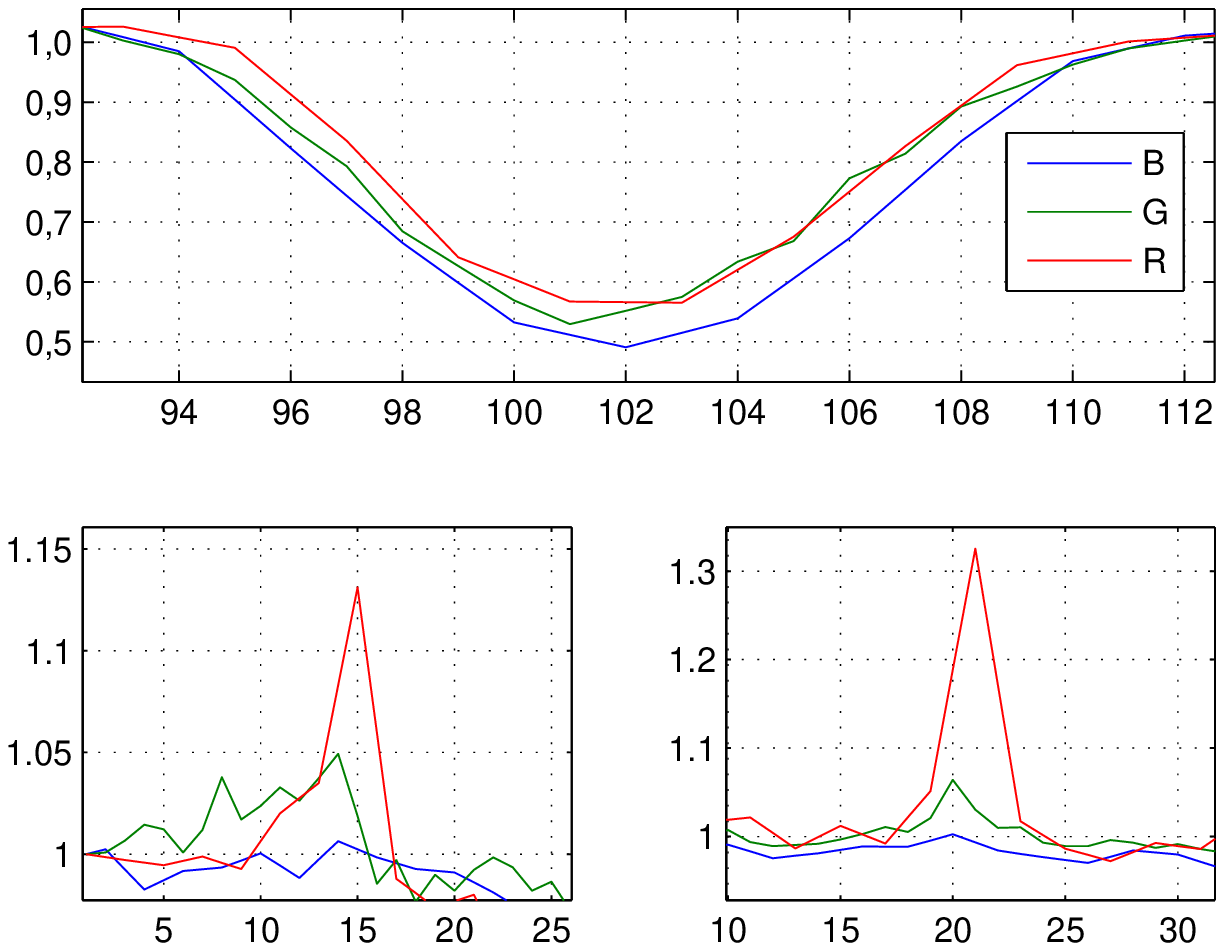,width = 0.9\linewidth} \caption{The color diagram gives distance of 6 km (top panel). RGB spectra of swifts (bottom panel).}\label{fig6}
\end{minipage}
\end{figure}

\begin{figure}[!h] 
\centering
\begin{minipage}[t]{.45\linewidth}
\centering
\epsfig{file = 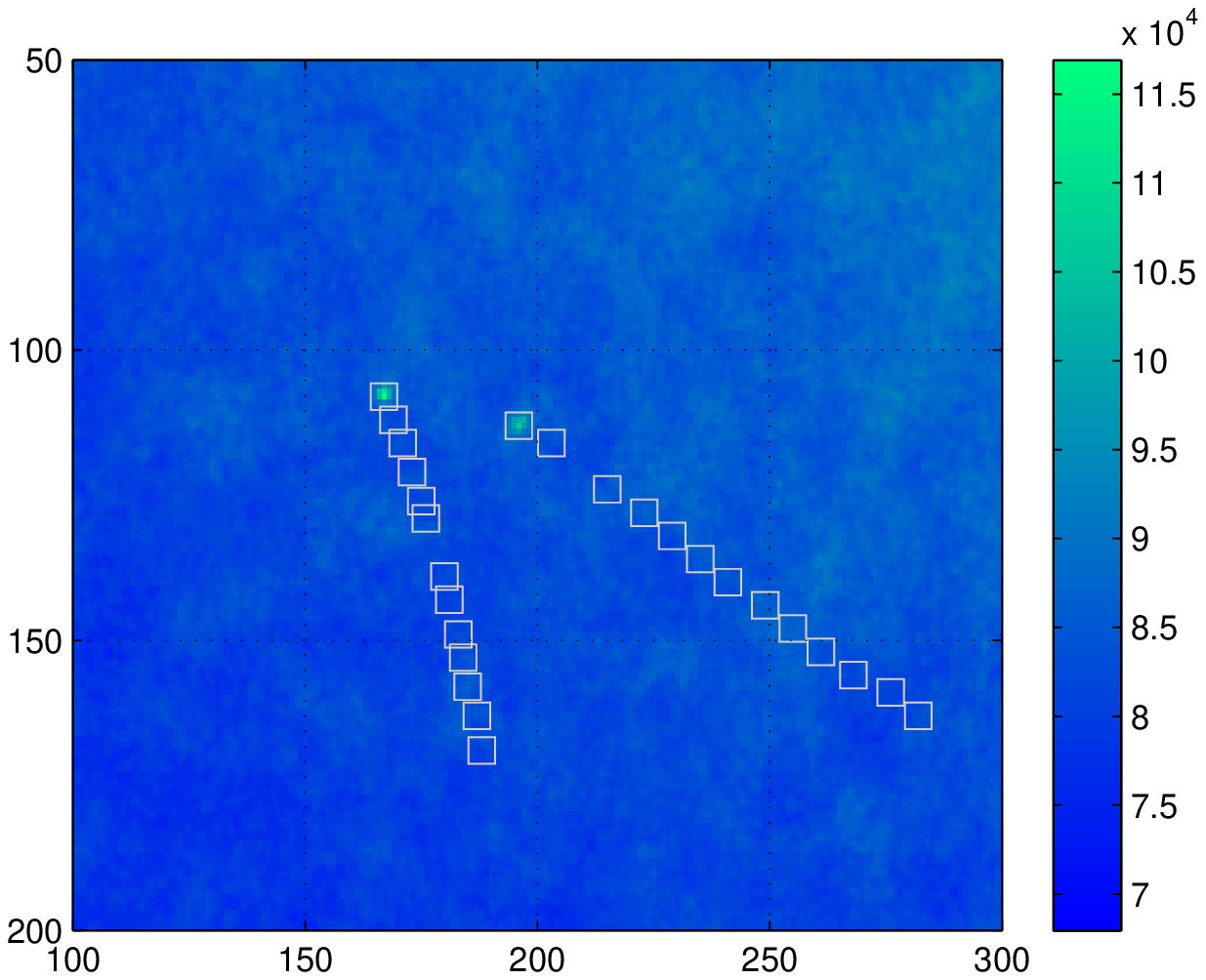,width = 0.9\linewidth} \caption{An image with two bright swifts with variable intensities.}\label{fig5}
\end{minipage}
\hfill
\begin{minipage}[t]{.45\linewidth}
\centering
\epsfig{file = 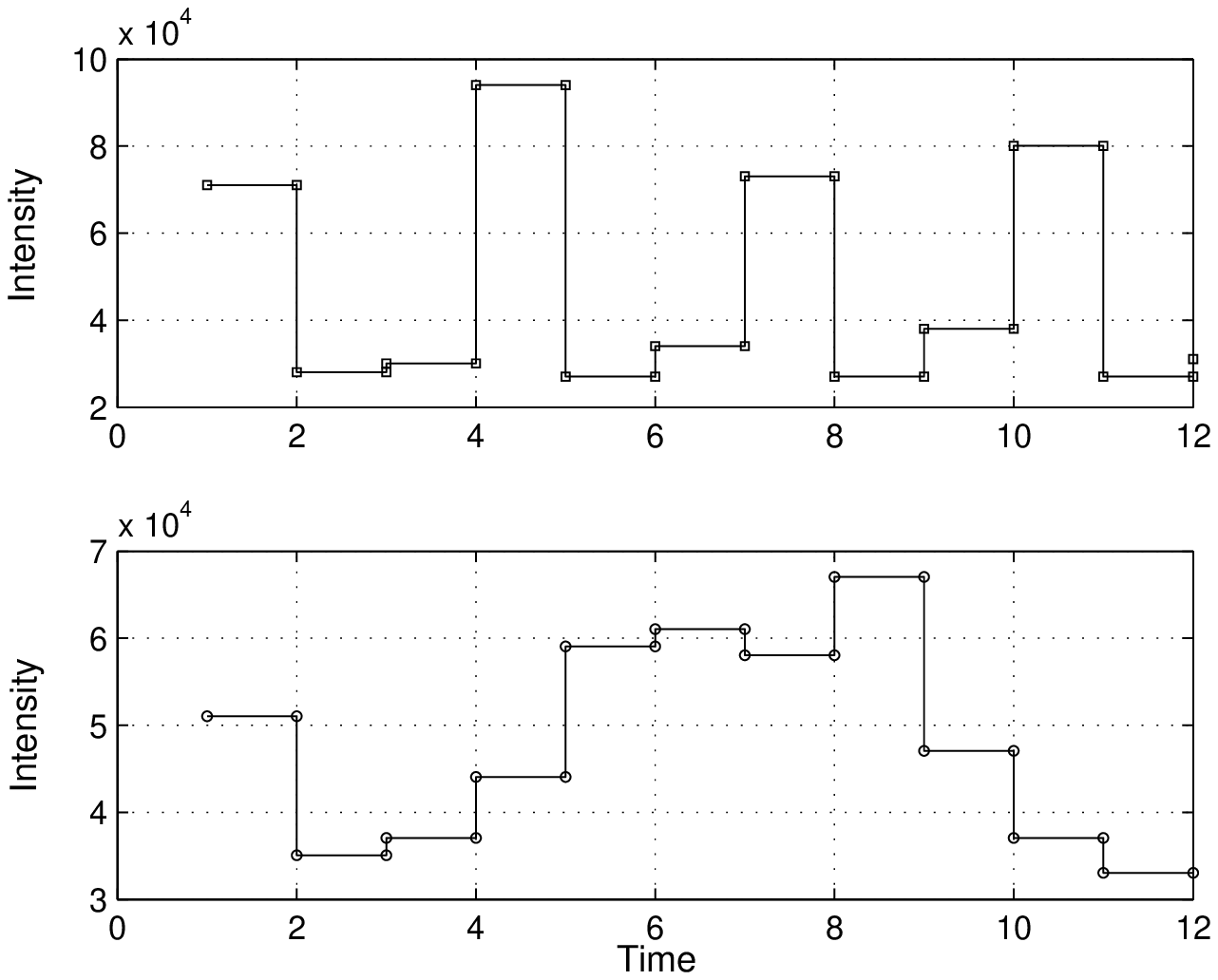,width = 0.9\linewidth} \caption{The light curves of two bright swifts with the sampling time of  20 ms.}\label{fig6}
\end{minipage}
\end{figure}

Fig. 15 shows an image with two bright swifts with variable intensities. Objects cross frame of 3 degrees with 50 frames per second with 1 ms exposure.  For 0.35 sec they demonstrate speed of 8 degrees per second. 

Fig. 16 shows the light curves of two bright swifts with the sample time of  20 ms. One swift demonstrates regular intensity variations of about 25 Hz. Another shows variations of about 10 Hz.

Figures 17 and 18 show UAPs over Kyiv. Objects cross frame of 2.2 degrees for 0.40 sec with 50 frames per second with 1 ms exposure. They demonstrate speed of 5.5 degrees per second.

Fig. 17 shows a composite image with the bright eagle and swift. It is obtained by dividing of two consecutive frames. We can see that objects are moving at different speeds.

Fig. 18 shows  an object called by us as "eagle". An object is in size of about 12.5 arc minutes, which indicates the final dimensions. Its contrast is about of 28\%.

If we assume that the "eagle" is at a distance of 1 km, its size will be about 6 meters, if at a distance of 4 km, then 25 meters. In the latter case, its speed will be about 380 m/s (about 1M).

Fig. 19 shows a composite image with the bright falcon, swift, and high-speed phantom. Figure presents a broad range of UAPs. We see them everywhere. We observe a significant number of objects whose nature is not clear.


Fig. 20 demonstrates the phantom crosses the image of the bright falcon. It is easy to see that the phantom is indeed an opaque body that shields the radiation of a bright object.

Fig. 21 demonstrates two-site observations of UAPs. It is necessary to synchronize two cameras with an accuracy of one millisecond. Shoot at a rate of at least 50 frames per second is needed. In a field of view of 5 degrees at a base of 120 km, objects above 1000 km can be detected.

An object against the background of the Moon was detected at zenith angle 56 degrees. Parallax about 5 degrees was evaluated. This allow us to evaluate distance equal to 1524 km, altitude 1174 km, and linear speed of 282 km/s.

Coincidence of 2-point light curves in Fig. 22 means: we observe the same object. Fig. 23 shows the light curve at a sampling rate of 125 Hz. The object flashes for one-hundredth of a second at an average of 20 times per second.



\begin{figure}[!h] 
\centering
\begin{minipage}[t]{.45\linewidth}
\centering
\epsfig{file = 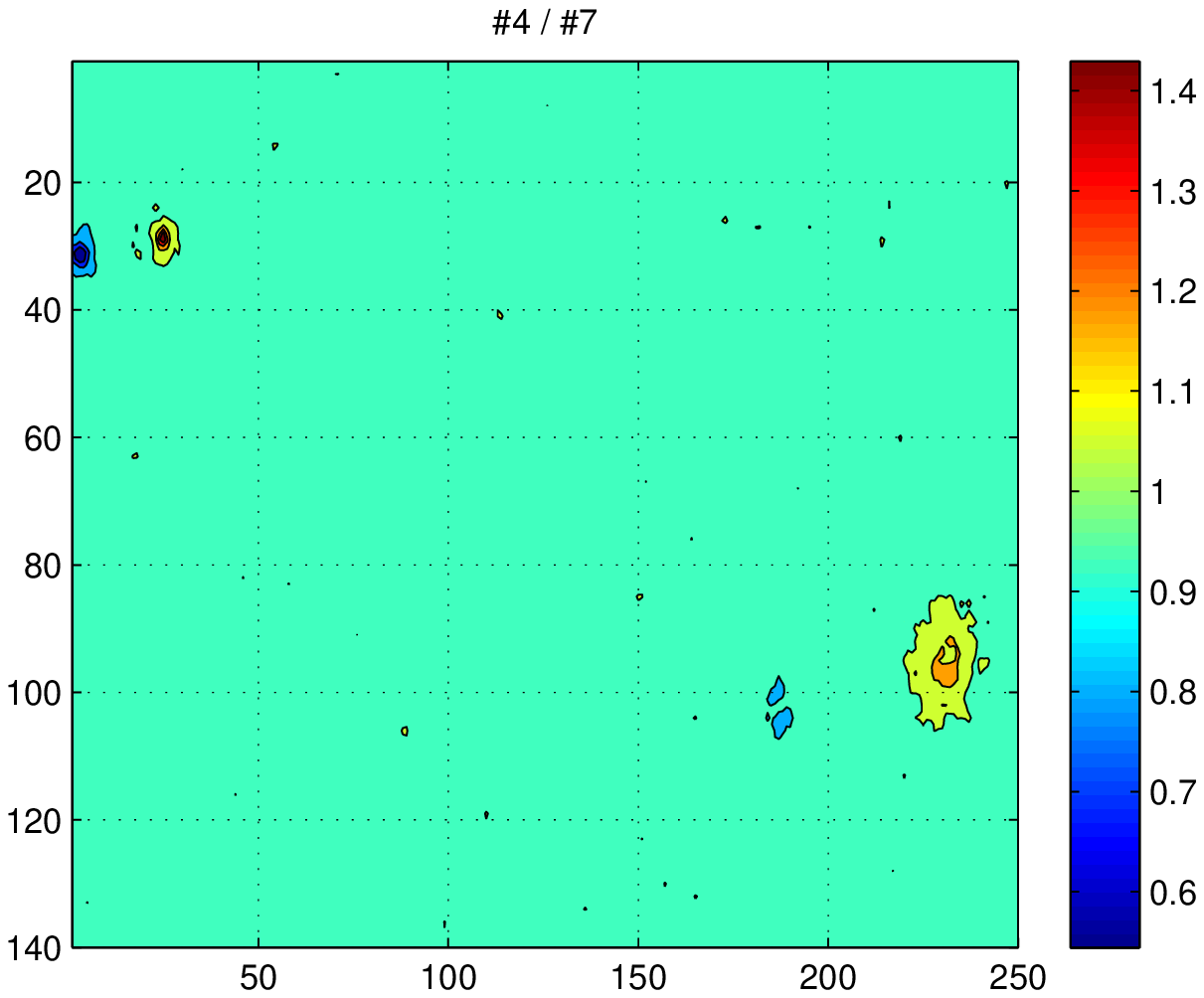,width = 0.9\linewidth} \caption{A composite image with the bright eagle, swift in the daytime sky over Kyiv.}\label{fig5}
\end{minipage}
\hfill
\begin{minipage}[t]{.45\linewidth} 
\centering
\epsfig{file = 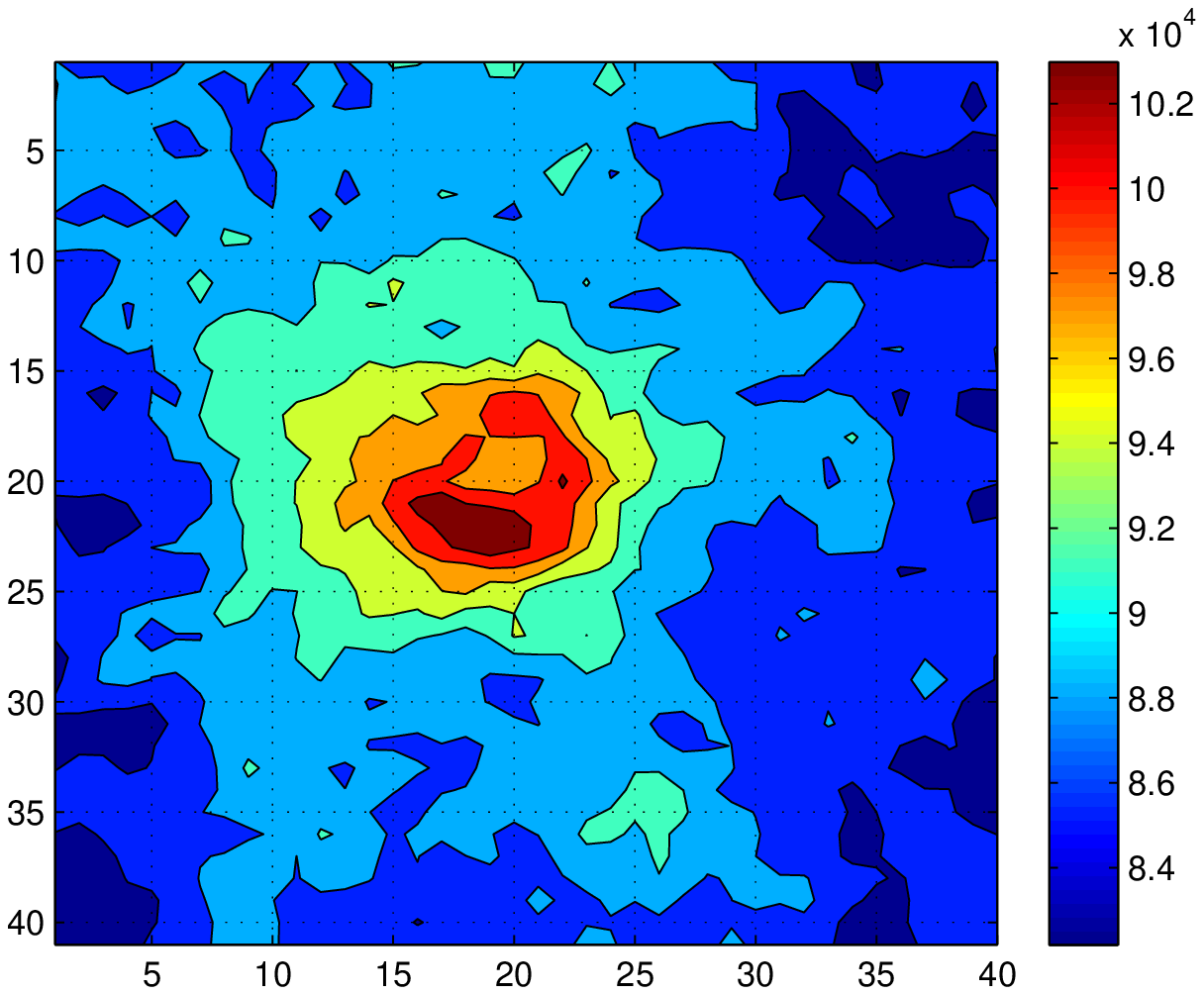,width = 0.9\linewidth} \caption{Structured image of an eagle in the sky over Kyiv.}\label{fig6}
\end{minipage}
\end{figure}

\begin{figure}[!h]
\centering
\begin{minipage}[t]{.45\linewidth}
\centering
\epsfig{file = 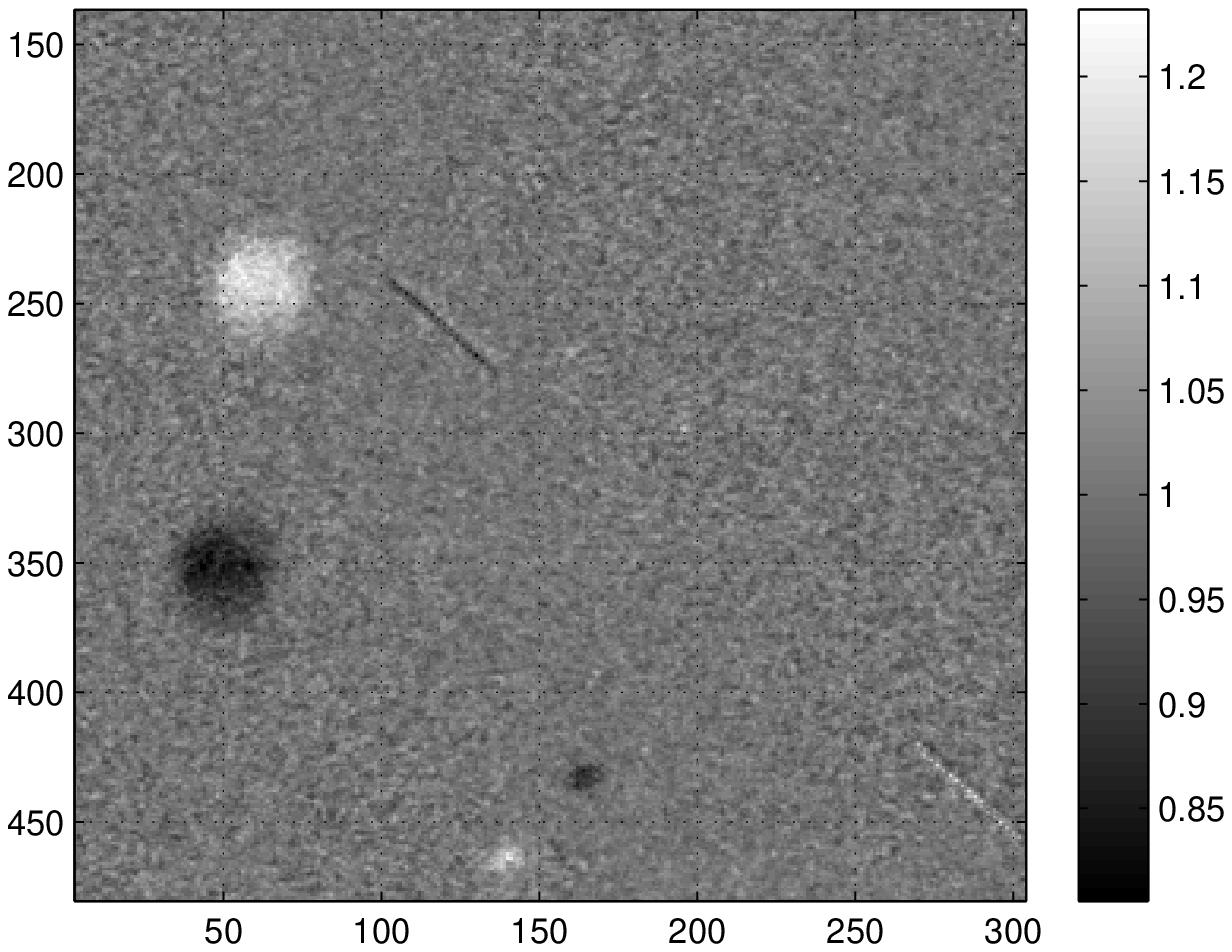,width = 0.9\linewidth} \caption{A composite image with the bright falcon, swift, and high-speed phantom.}\label{fig5}
\end{minipage}
\hfill
\begin{minipage}[t]{.45\linewidth}
\centering
\epsfig{file = 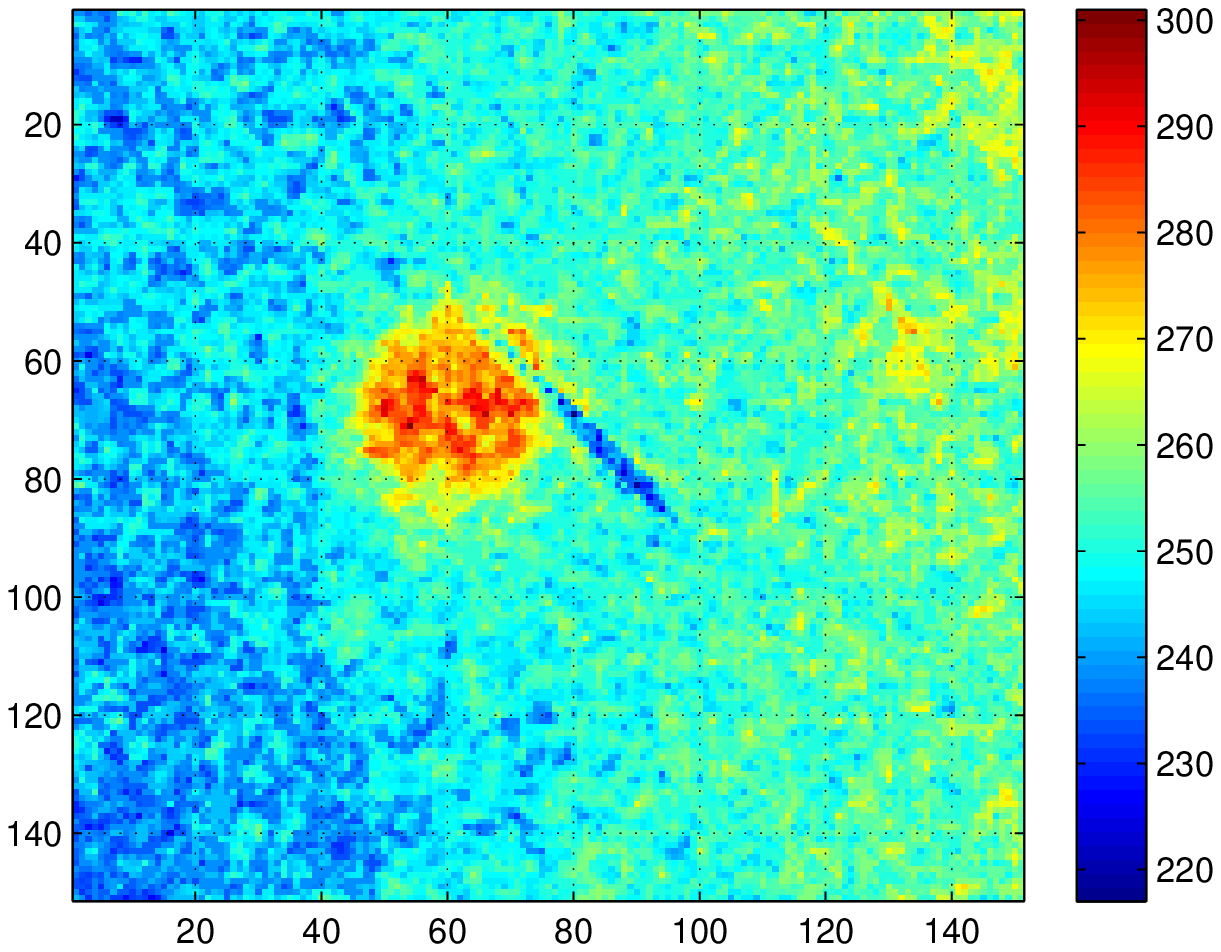,width = 0.9\linewidth} \caption{The phantom crosses the image of the falcon.}\label{fig6}
\end{minipage}
\end{figure}

\begin{figure}[h]
\centering
\resizebox{0.45\hsize}{!}{\includegraphics[angle=000]{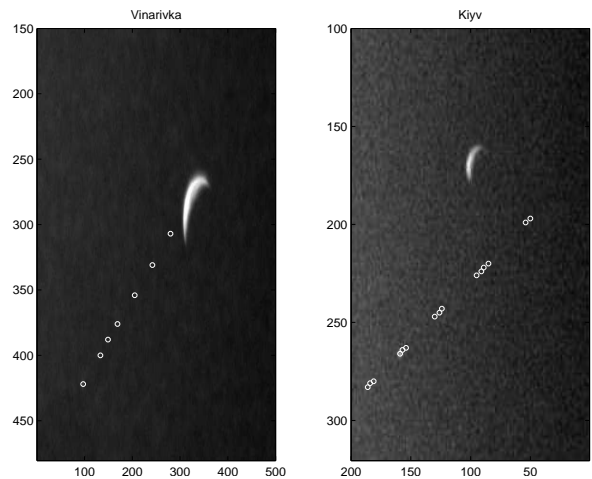}}%
\caption{Two-site observations.} \label{figure: HAT_P_1_14qz.eps}
\end{figure}

\begin{figure}[!h]
\centering
\begin{minipage}[t]{.45\linewidth}
\centering
\epsfig{file = 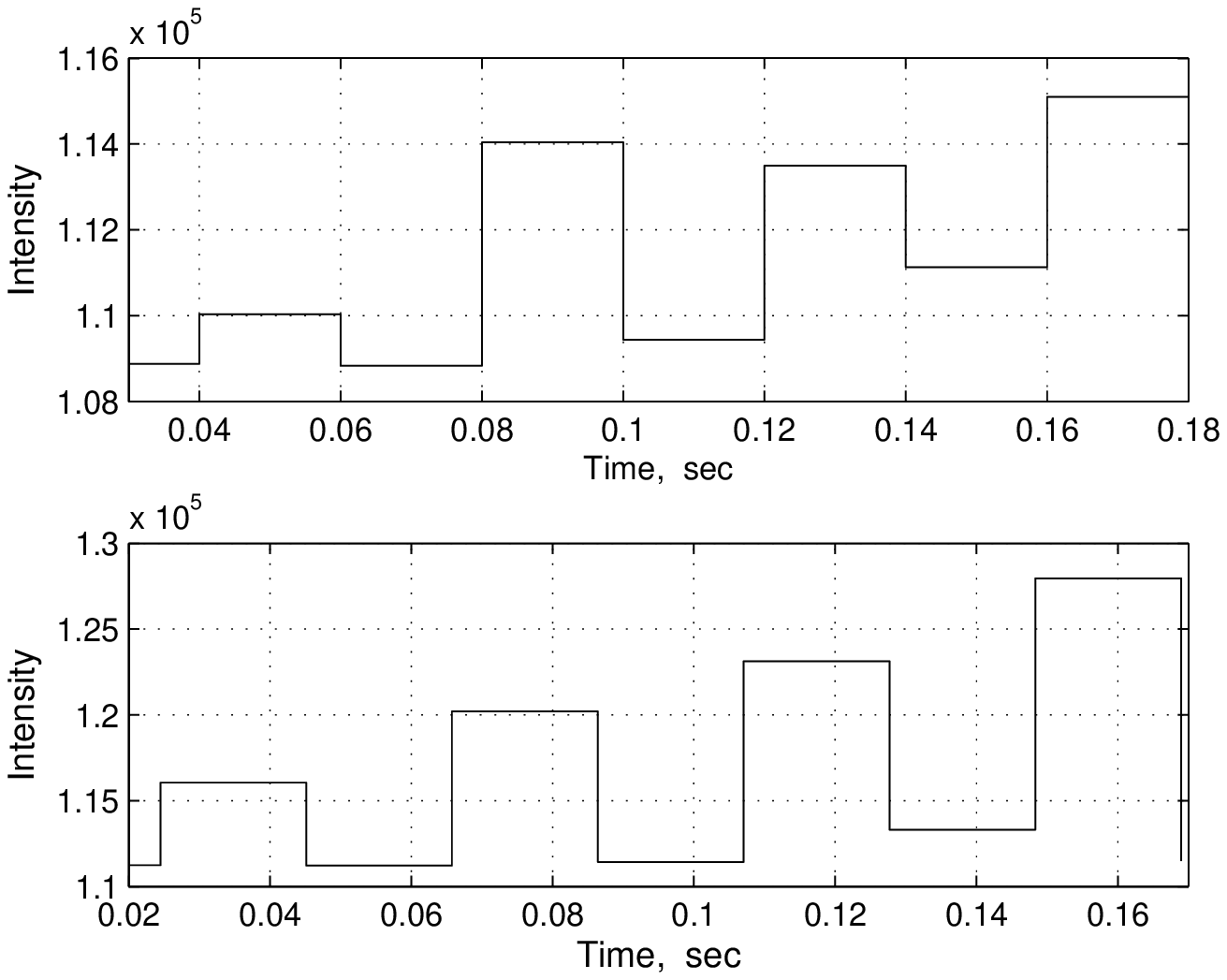,width = 0.9\linewidth} \caption{Two-point object light curves.}\label{fig5}
\end{minipage}
\hfill
\begin{minipage}[t]{.45\linewidth}
\centering
\epsfig{file = 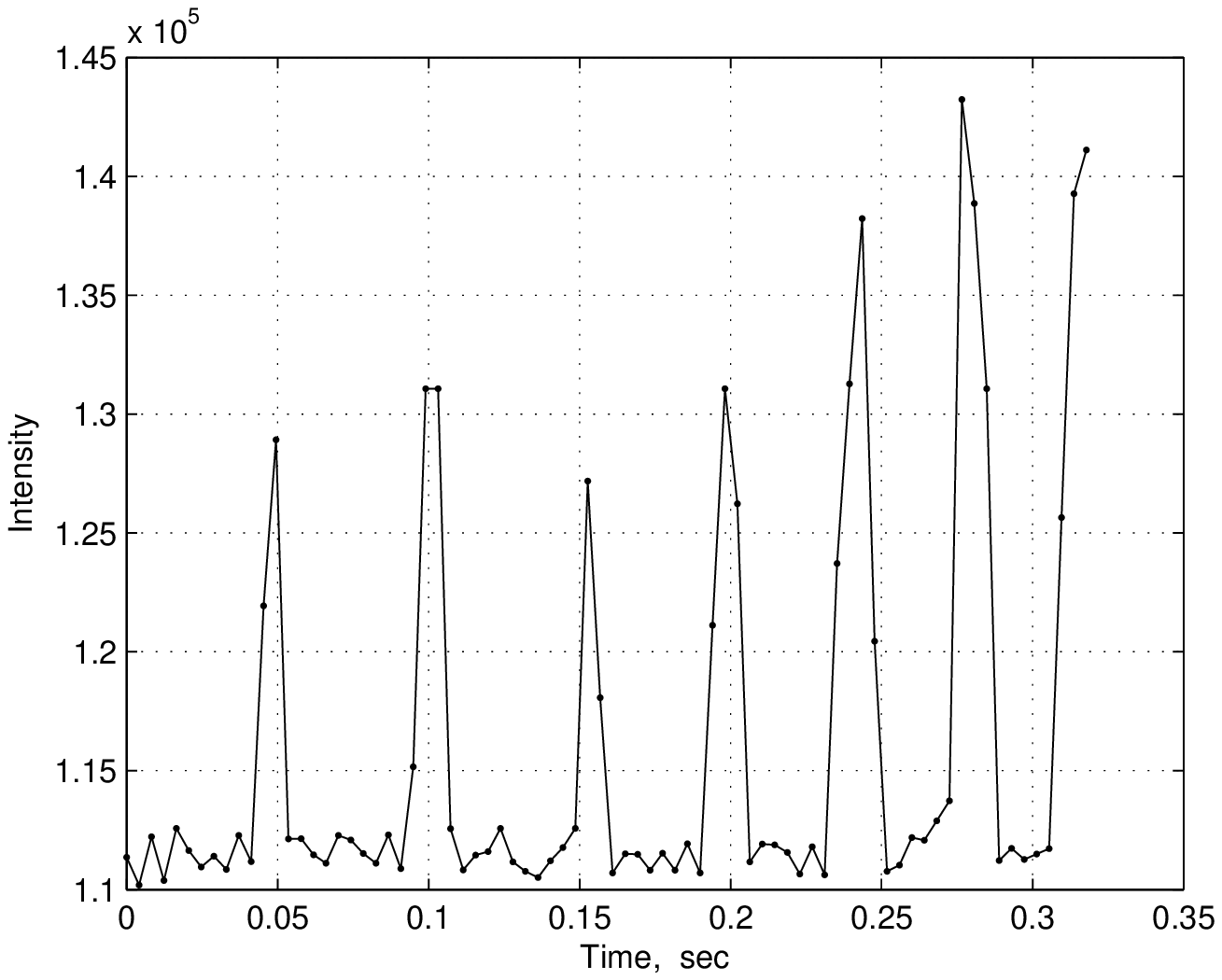,width = 0.9\linewidth} \caption{The light curve at a sampling rate of 125 Hz.}\label{fig6}
\end{minipage}
\end{figure}

\vspace*{1ex}

\vspace*{1ex}


\section*{\sc Conclusions}

The Main Astronomical Observatory of NAS of Ukraine conducts a study of UAP.  We used two meteor stations installed in Kyiv and in the Vinarivka village in the south of the Kyiv region. 

Observations were performed with colour video cameras in the daytime sky. A special observation technique had developed for detecting and evaluating UAP characteristics.

There are two types of UAP, conventionally called Cosmics, and Phantoms. Cosmics are luminous objects, brighter than the background of the sky. Phantoms are dark objects, with contrast from several to about 50 per cent.

We observed a broad range of UAPs everywhere. We state a significant number of objects whose nature is not clear.

Flights of single, group and squadrons of the ships were detected, moving at speeds from 3 to 15 degrees per second. Some bright objects exhibit regular brightness variability in the range of 10 - 20 Hz.

Two-site observations of UAPs at a base of 120 km with two synchronised cameras allowed the detection of a  variable object, at an altitude of 1170 km. It flashes for one hundredth of a second at an average of 20 Hz.

Phantom shows the colur characteristics inherent in an object with zero albedos. We see an object because it shields radiation due to Rayleigh scattering. An object contrast made it possible to estimate the distance using colorimetric methods.

Phantoms are observed in the troposphere at distances up to 10 - 12 km. We estimate their size from 3 to 12 meters and speeds up to 15 km/s.


\end{document}